# A Digital Guitar Tuner

Mary Lourde R., Anjali Kuppayil Saji,
*Department of Electrical & Electronics Engineering*
*BITS-Pilani, Dubai, UAE*
Email Id.: marylrd@yahoo.com, Anjali.saji@gmail.com

*Abstract* - **The objective of this paper is to understand the critical parameters that need to be addressed while designing a guitar tuner. The focus of the design lies in developing a suitable algorithm to accurately detect the fundamental frequency of a plucked guitar string from its frequency spectrum. A user-friendly graphical interface is developed using Matlab to allow any user to easily tune his guitar using the developed program.**

## I. INTRODUCTION

In order to play beautiful music, a musician needs to have perfectly tuned instruments. As any novice musician knows, tuning an instrument can be difficult for the untrained ear. This paper discusses a method to design a digital guitar tuner. The tuner is designed using the filtering, measuring and analyzing capabilities of Matlab. Compared to tuning by ear, where a certain amount of guess work is involved in deciding how much to tighten/loosen a string, the Matlab based tuner is able to give accurate instructions so that tuning can be achieved quicker.

A guitar note is not made up of a single frequency. It consists of a number of harmonics as well. The difference in the harmonics is what makes a guitar and a violin sound different even while playing the same note.[3] Once a sample note is played on the guitar, its fundamental frequency has to be calculated. A major part of the project is dedicated towards recovering the fundamental frequency from the frequency spectrum of the sample note played on a guitar.

Another limitation that has to be overcome is the low resolution of the frequency spectrum. The human ear is insensitive to variations smaller than +/- 0.5 Hz. In order to obtain real time data, sampling interval has to be kept as low as possible. However, this reduces the resolution of the spectrum and hence affects the accuracy of frequency estimation. Different methods to overcome this are analyzed and a suitable method is applied to meet the requirement.

## II. FINDING THE FUNDAMENTAL FREQUENCY OF PARTICULAR GUITAR STRING

The first phase of the project deals with finding the frequency of a played guitar note. In order to do this, the string is plucked once and the sound is sampled into an array variable. Filters maybe used to eliminate noise as well as unwanted frequencies above/below the expected range. The signal is converted to frequency domain using fast Fourier transform. However, a computer is able to represent signals using only discrete values. Therefore the signal that is converted to frequency domain is plotted at equally spaced discrete positions. These equally spaced positions are known as bins. The smaller the distance between the bins, the more accurate the signal will be.

The frequency resolution (i.e. the distance between the bins) is calculated using the following formula: [1]

Resolution = Sampling Rate / No of Samples    (1)

Using a sampling rate (Fs) = 8000 Hz and a sampling time of 0.5 seconds
  Resolution = 8000/ (0.5*8000) = 2 Hz

However, this resolution is not good enough as a trained human ear can distinguish differences as small as 0.5 Hz. Therefore, we need to increase the resolution. A suitable method needs to be used to improve the resolution of the frequency bins in order to achieve accurate estimation of the frequency of the signal. Another simpler solution to increase frequency resolution of the Fast Fourier transform would be to increase the duration of sampling. This would have some drawback in terms of increasing data acquisition time.

Another limiting factor is that we cannot increase the sampling time by a large amount as we can only get a maximum of 2 to 3 sec of real data from a single guitar pluck.

Setting the sampling time as 2 sec and repeating the above calculations,
Using a sampling rate (Fs) = 8000 Hz
  Resolution = 8000/ (2*8000) = 0.5 Hz
This resolution is suitable for our purpose.

The fundamental frequency of the string is taken as the frequency bin corresponding to the maximum magnitude in the Fast Fourier Transform(FFT).

## A. Filtering the signal using an FIR filter

The Fast Fourier Transform of the original signal is shown below in Figure 1. We can see that there is a large amount of noise in the signal. This may adversely affect the calculations required for frequency detection. In order to overcome this, an appropriate filter has to be applied to the signal.

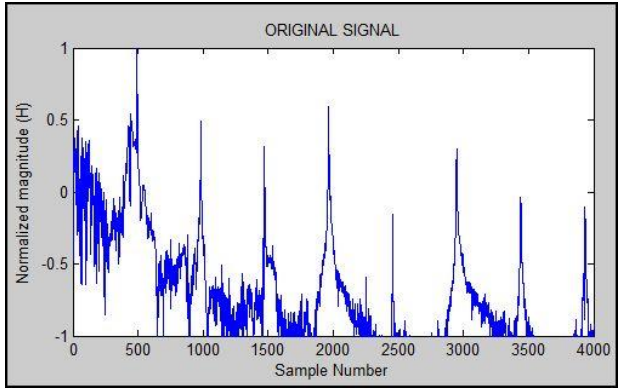

Figure 1: Normalized frequency spectrum of the open B string played on the guitar

It is proposed to apply a band pass filter to the signal that is created using a hamming window. The hamming window provides a smooth transition and provides linear phase characteristics. It provides an attenuation greater than 50 db for the side lobes which is suitable for our requirement.

The maximum expected fundamental frequency from the open string of a guitar is about 440 Hz.[2] The harmonics from the string will occur at multiples of the fundamental frequency. Our algorithm requires the use of at least 3 harmonics from the signal. Therefore we require the filter to pass frequencies up to approximately 440*3 = 1320 Hz. The minimum expected frequency from the open string of a guitar is about 75 Hz. The characteristics of the hamming window filter used are shown in Figure 2. Note that in this graph the x axis corresponds to the frequency and not the sample number.

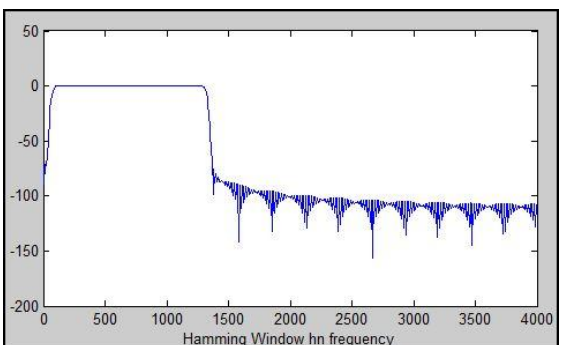

Figure 2: Hamming window

A transition bandwidth of 25 Hz is determined to provide sufficient attenuation so as remove noise especially those at the lower boundary of the pass band. The cut-off frequencies of the band pass filter are placed at 75 and 1320 Hz. Since the frequency resolution is 0.5 (sampling time is 2sec), we multiply the frequencies by 2 to get the corresponding sample numbers (i.e. 150 and 2640 Hz respectively). The FFT of the signal after filtering is shown below in Figure 3.

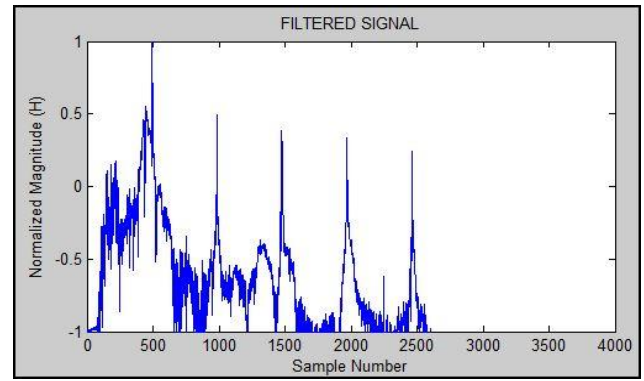

Figure 3: Fast Fourier Transform of Signal after filtering

## B. Using Harmonic Data to detect frequency

In order to develop more accurate frequency detection, we take advantage of the harmonic data present in the signal. When a guitar string is plucked, it vibrates not only at the fundamental frequency but also at frequencies that are multiples of the fundamental frequency. These multiples of the fundamental frequency are known as harmonics. [5]

Therefore, if we down sample the obtained signal by a factor of 2, the 1st harmonic would lie at the position (sample number) corresponding to the fundamental frequency of the original signal. Down sampling the frequency by a factor of 3, the 2nd harmonic would lie at the position (sample number) corresponding to the fundamental frequency of the original signal and so on. Hence, by analysis, we find that adding or multiplying these signals together would produce the maximum peak at the fundamental frequency of the signal [5]. This can be easily understood from the illustration shown below Figure 4.

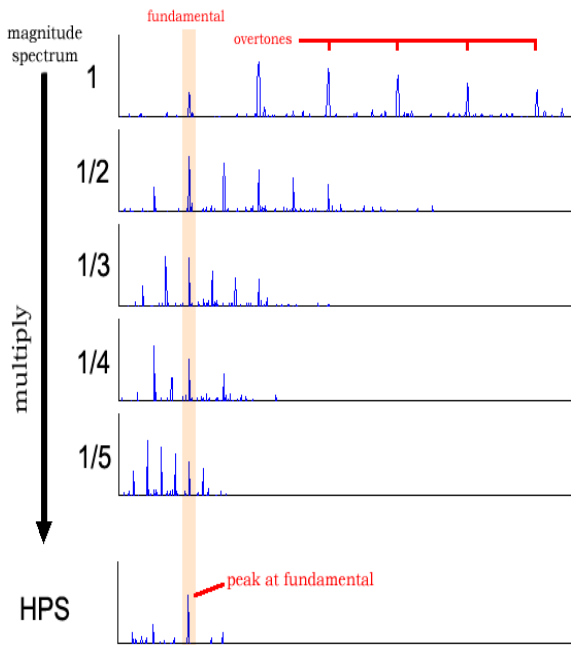

Figure 4: Harmonic Product Spectrum [5]

Another advantage of using harmonic data present in the signal is that it gives accurate results. In some cases, it may not be the fundamental frequency that has the maximum magnitude in the Fourier transform of the signal. In some cases, the 1st harmonic has greater amplitude. This would lead to inaccurate results if harmonic data was not used. The specialty about the harmonic frequencies is that they occur at intervals equal to the fundamental frequency. Hence by performing Harmonic addition we ensure that only the fundamental frequency gets amplified, and hence frequency detection is accurate. This detection algorithm was tested in Matlab and various parameters were adjusted to arrive at the optimal result.

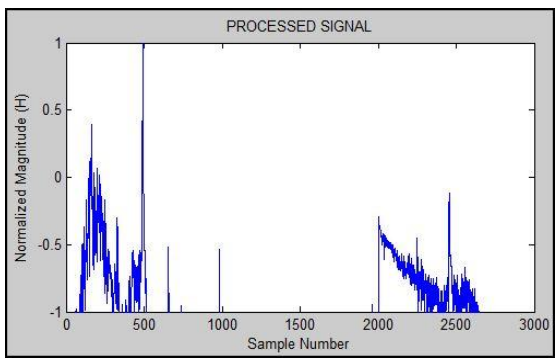

Figure 5: The Fast Fourier Transform of the signal after adding the down sampled components

In the final implementation, the signal is down sampled twice (that is by a factor of 2 and by a factor of 3). The down sampled signals are then zero padded at the end in order to retain the 4000 sample length of the original signal. This allows us to add the down sampled signals to the original signal and hence obtain the harmonic addition spectrum.

The FFT of the signal after adding the down sampled components to the original signal are shown in Figure 5. We see that a clear peak emerges at the fundamental frequency.

### III. COMPARING THE FREQUENCY WITH THE STANDARD FREQUENCY

In the second phase, the frequency of the played note is compared with standard value that is desired for that particular note. An octave is divided into 12 equally spaced notes. The distance between each of the consecutive notes is known as a semi-tone. Every twelve notes that fill one octave are then represented by the symbols: C, $C^{\#}$ or $D^{b}$, D, $D^{\#}$ or $E^{b}$, E, F, $F^{\#}$ or $G^{b}$, G, $G^{\#}$ or $A^{b}$, A, $A^{\#}$ or $B^{b}$, B

The location of note A4 is defined as 440 Hz. The frequency of the remaining notes can be calculated using the formula given below.

$$\text{Frequency} = 440 * 2^{(n/12)} \quad (2)$$

where n is the number of semitones above A4

Using this formula, the frequency of the remaining notes are calculated and tabulated in the following table. (Table I). [3]

TABLE I
STANDARD FREQUENCIES FOR GUITAR STRINGS

| Open Guitar Strings | Key Number | Standard Frequency (Hz) |
|---|---|---|
| E | E2 | 82.4 |
| A | A2 | 110.0 |
| D | D3 | 146.8 |
| G | G3 | 196.0 |
| B | B3 | 246.9 |
| E | E4 | 329.6 |
|   | A4 | 440.0 |

First the detection process was tested for the E4 string. Later the similar process was replicated for all the other strings. The user is allowed to choose which string he wants to tune and the appropriate frequencies are used for analysis.

During the initial testing, it was observed that it was more difficult to accurately detect the fundamental frequency of the lower frequency strings. One of the reasons for this was attributed to the high noise content at the low frequency range. This was compensated by using an appropriate filter as described above.

On further research, some interesting observations were made. Musical notes are spaced at logarithmic intervals. This implies that notes at lower frequencies are spaced closer together compared to those at higher frequencies. However, the frequency bins used during Fast Fourier Transform are equi-spaced on a linear scale. Therefore at lower frequencies, it is more difficult to distinguish between the different notes as more than one note may map to the same frequency bin. [4]

Different methods have been proposed to overcome this problem. One method involves developing an algorithm to transform the signal to frequency domain using a logarithmic scale instead of linear scale. This would space the frequency bins at logarithmic distances and hence overcome the problem. However, for this project, this slight difference in the detected values has been considered inconsequential.

## IV. DETERMINING THE AMOUNT OF TURN TO BE GIVEN TO TUNING KNOB OF THE GUITAR

The change in frequency of a string per quarter turn (or per degree of turn) of the tuning knob was found by taking samples. Based on the data the *average change* in frequency per quarter turn (or per degree turn) will be calculated and used as a reference for tuning any guitar. Figure 6 shows the location of the tuning pegs for the different strings of an acoustic guitar.

Based on the difference between the actual frequency and desired frequency of the string, the amount of turn to be given to the string will be displayed. For complete automation of the guitar, further work can explore the possibility of using a dc stepper motor to automatically turn the string by the required amount.

### A. Data Collection

The frequency of the string was first tested and then another sample taken after turning the tuning peg clockwise by 180. The tuning peg was turned anti-clockwise by 180 and the next sample was taken. This process was repeated 3 times for each string. The following were the data obtained by testing each of the strings (Table II).

From the table we see that there is negligible difference between tightening and loosening in a particular string. However, there is a large difference in the values for different strings. This is because material for each string is made slightly differently depending of the frequency range within which it is expected to be used.

From the analysis of the data collected, the value for tightening is taken to be same as that for loosening the string. Using this approximation, the change in frequency per degree of turn is calculated for each string by dividing the obtained value by the number of degrees of turn. i.e. 180 degrees. The values are tabulated in Table III.

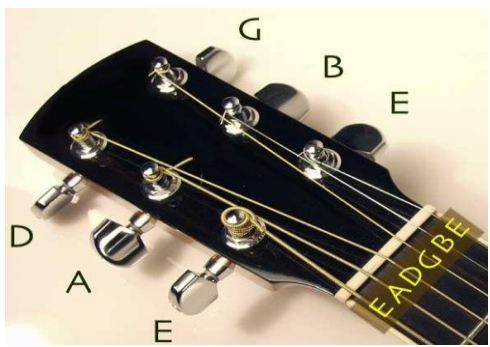

Figure 6: Tuning Pegs of an Acoustic Guitar

TABLE II
FREQUENCY CHANGE FOR DEGREES OF TUNING PEG TURN

| STRING | E (STRING 6) | | |
|---|---|---|---|
| | Change in frequency | | |
| | Test 1 | Test 2 | Test 3 |
| Turning clockwise 180° | + 4 Hz | + 4 Hz | + 4Hz |
| Turning anticlockwise 180° | - 4 Hz | - 4 Hz | - 4 Hz |
| **STRING** | **A (STRING 5)** | | |
| | Change in Frequency | | |
| | Test 1 | Test 2 | Test 3 |
| Turning clockwise 180° | + 4.5 Hz | + 4.5 Hz | + 4.5 Hz |
| Turning anticlockwise 180° | - 4.5 Hz | - 4.5 Hz | - 4.5 Hz |
| **STRING** | **D (STRING 4)** | | |
| | Change in Frequency | | |
| | Test 1 | Test 2 | Test 3 |
| Turning clockwise 180° | + 5.5 Hz | + 5 Hz | + 5 Hz |
| Turning anticlockwise 180° | - 5 Hz | - 5 Hz | - 5 Hz |
| **STRING** | **G (STRING 3)** | | |
| | Change in Frequency | | |
| | Test 1 | Test 2 | Test 3 |
| Turning clockwise 180° | + 12 Hz | + 12 Hz | + 12 Hz |
| Turning anticlockwise 180° | - 12 Hz | - 12 Hz | - 12 Hz |
| **STRING** | **B (STRING 2)** | | |
| | Change in Frequency | | |
| | Test 1 | Test 2 | Test 3 |
| Turning clockwise 180° | + 13.5 Hz | + 13.5 Hz | + 13.5 Hz |
| Turning anticlockwise 180° | - 12.5 Hz | - 13.5 Hz | - 13.5 Hz |
| **STRING** | **E (STRING 1)** | | |
| | Change in Frequency | | |
| | Test 1 | Test 2 | Test 3 |
| Turning clockwise 180° | + 16 Hz | + 15.5 Hz | + 15 Hz |
| Turning anticlockwise 180° | - 15 Hz | - 15 Hz | - 15 Hz |

TABLE III
CHANGE IN FREQUENCY PER DEGREE OF TURN

| STRING | Change in Frequency per degree turn | |
|---|---|---|
| | Calculation | Final Value (Hz/degree) |
| E (STRING 6) | 4/180 | 0.022 |
| A (STRING 5) | 4.5/180 | 0.025 |
| D (STRING 4) | 5/180 | 0.028 |
| G (STRING 3) | 12/180 | 0.067 |
| B (STRING 2) | 13.5/180 | 0.075 |
| E (STRING 1) | 15/180 | 0.083 |

These values are used to estimate the number of degrees of turn required to bring an out of tune string to the proper frequency. This is done in the following way.

The difference between the desired value of the frequency is obtained from Table I. The difference between this value and the current value of the string (obtained by analysis of the 2 sec input) is multiplied with the value for the corresponding string in Table III. Based on whether the difference was positive or negative, the appropriate action is suggested to the user.

## V. DEVELOPING A GRAPHICAL USER INTERFACE (GUI)

A GUI has been developed to provide a user friendly interface for the whole application. This was done using Matlab GUI builder. The features of the interface are

(i) Allowing the user to select which of the six strings is to be tuned.
(ii) Displaying the Frequency plot of the recorded signal to ensure that the signal has been recorded properly.
(iii) Displaying the Frequency plot of the processed signal.
(iv) A button called 'Test string' to start testing the selected string.
(v) A display board which instructs when to play the string and when to stop.
(vi) The correct frequency that each string should have is displayed alongside the string name.
(vii) Display boards for each of the strings show the detected frequency as a result of processing the signal. It also suggests the approximate angle by which an out of tuned string should be turned in order to bring it to the correct frequency.
(iv) Tuning knobs for each of the strings which display the amount of degrees by which the tuning peg should be turned for tuning the string.

Implementation of these features involved writing code for the callback functions for each of the controls. The properties of each control were set to the desired values and code was written to change certain values based on user actions and results from the signal processing.

Some of the controls used to prepare the final graphical interface included the following:

(i) Push Buttons – 6 for selecting one of the six strings and 1 for the Test String button
(ii) Edit Text boxes – 1 box for each string to display Play/Stop instructions as well as to suggest appropriate action to tune the string.
(iii) ActiveX Control Tuning knobs – To show the angle of turn required to tune each of the six strings.
(iv) Panels to organize the layout of the interface.

The layout of the GUI was optimized for maximum ease of use. Care was also taken to display the maximum information possible within the available area. A snapshot of the final layout is shown below in Figure 7.

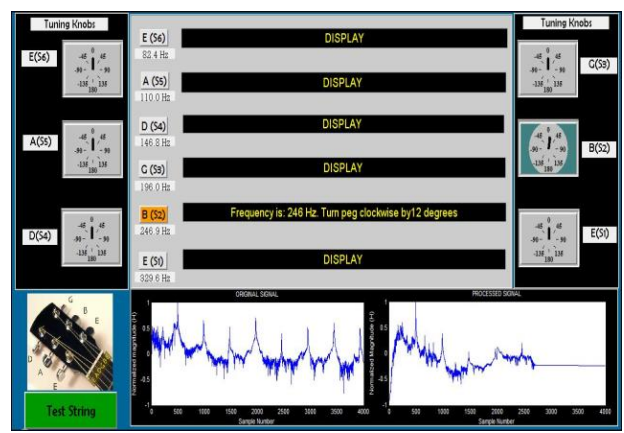

Figure 7: Layout of the GUI

A particular string can be selected by clicking the button with the corresponding name. The button then becomes highlighted in orange. The 'TEST STRING' button can then be pressed once the user is ready to struck the required string. The display board gives instruction when the player has to start playing the string. After 2 seconds the display board instructs the player to stop. The program then performs the required analysis. The result of the analysis and the suggested action is then displayed on the corresponding display board. Simultaneously, the associated tuning knob turns to show the approximate degree by which the string has to be tightened or loosened to achieve perfect tuning.

The layout provides the user with graphs of the original signal as well as the processed signal. This is helpful for users who want to analyze the signals further. A user who just needs to tune his instruments can directly read the values and suggested actions from the display boards and tuning knobs.

## VI. SIMULATION OF AUTOMATIC TUNING OF THE GUITAR

In order to achieve complete automation of guitar tuning, the next logical step of this project would be to design a circuit to continue tightening or loosening a virtual guitar string until the frequency is within acceptable range of the standard value. Once the angle by which the tuning peg has to be turned is determined, a proportional voltage has to be supplied to a dc stepper motor that is attached to a clamp on the tuning peg.

Though simulation of this process was attempted using Matlab, the hardware implementation of automating the guitar tuning is beyond the scope of this paper. However, future work on this project can be focused in this direction.

## CONCLUSION

Design of a digital guitar tuner is discussed in this paper. The describes the details on the factors involved in tuning of a guitar. The important factor being the fundamental frequency of the notes, the methods to find the same is discussed using Harmonic Product Spectrum. The factors that affect fundamental frequency detection such as sampling frequency, sampling duration, background noise and the logarithmic nature of signal intervals in musical notes etc are discussed in detail. The concept that the harmonic nature of guitar notes could be used to accurately determine the fundamental frequency was instrumental in simplifying the frequency detection algorithm.

Developing the graphical user interface made the application much more practical in that even those who do not know how to use MatLab can still easily use the application to tune a guitar.

This project can be extended to (Hardware implementation) include automatic tuning of the guitar strings by using a motor to turn the tuning pegs by a voltage proportional to the amount of turn required.

## ACKNOWLEDGMENT

Ms. Anjali would like to express her sincere gratitude to Dr. Ramachandran, Director, BITS-Pilani, Dubai for the opportunity to undertake this project which has widened her understanding and sparked the interest in the field of digital signal processing. She is also thankful to Dr. Mary Lourde, Head of Department, Electrical and Electronics Engineering, for her constant guidance and support throughout the course of the project. Her valuable insights and attention to the details helped make this project a success.

## REFERENCES

[1] Gorrell, L. & R. S, "Automatic Guitar Tuner", Senior Design Project, Department of Electrical and Computer Engineering, University of Illinois at Urbana-Champaign, December 2002.

[2] Brain .M, "How Acoustic Guitars Work." , HowStuffWorks.com, September 2008. http://entertainment.howstuffworks.com/guitar.htm

[3] "Guitar Tuning", http://en.wikipedia.org/wiki/Guitar, February 2009

[4] Ilan L., "Automatic Guitar Tuner", Thesis paper, School of Information Technology and Electrical Engineering, University of Queensland

[5] "Harmonic Spectrum", June 2006 http://sv.mazurka.org.uk/MzHarmonicSpectrum/